\documentclass[aps,prb,twocolumn,superscriptaddress]{revtex4}
\usepackage{graphicx}
\usepackage{amsmath}
\usepackage{amssymb}
\usepackage{epstopdf}

\begin{document}

\title{Quasielastic neutron scattering of hydrated BaZr$_{0.90}$$A$$_{0.10}$O$_{2.95}$ ($A$ = Y and Sc)}

\author{Maths Karlsson}
\email[]{maths@fy.chalmers.se}
\affiliation{Department of Applied Physics,
             Chalmers University of Technology,
             SE-412 96 G{\"o}teborg, Sweden}
\author{Aleksandar~Matic}
\affiliation{Department of Applied Physics,
             Chalmers University of Technology,
             SE-412 96 G{\"o}teborg, Sweden}
\author{Dennis Engberg}
\affiliation{Department of Applied Physics,
             Chalmers University of Technology,
             SE-412 96 G{\"o}teborg, Sweden}
\author{M{\aa}rten E. Bj\"{o}rketun}
\affiliation{Department of Applied Physics,
             Chalmers University of Technology,
            SE-412 96 G{\"o}teborg, Sweden}
\affiliation{Center for Atomic-scale Materials Design,
	     Department of Physics,
             Technical University of Denmark,
             DK-2800 Lyngby, Denmark}	    
\author{Michael M. Koza}
\affiliation{Institut Laue-Langevin,
             F-38042 Grenoble CEDEX, France}
\author{Istaq Ahmed}
\affiliation{Department of Chemical and Biological Engineering, Chalmers University of Technology, SE-412 96 G{\"o}teborg, Sweden}
\author{G\"{o}ran Wahnstr\"{o}m}
\affiliation{Department of Applied Physics,
             Chalmers University of Technology,
             SE-412 96 G{\"o}teborg, Sweden}
\author{Pedro Berastegui}
\affiliation{Department of Inorganic Chemistry, Arrhenius Laboratory, Stockholm University, SE-106 91 Stockholm, Sweden}
\author{Lars B{\"o}rjesson}
\affiliation{Department of Applied Physics,
            Chalmers University of Technology,
             SE-412 96 G{\"o}teborg, Sweden}
\author{Sten Eriksson}
\affiliation{Department of Chemical and Biological Engineering, Chalmers University of Technology, SE-412 96 G{\"o}teborg, Sweden}

\date{\today}
\begin{abstract}

Proton motions in hydrated proton conducting perovskites BaZr$_{0.9}$$A$$_{0.10}$O$_{2.95}$ ($A$ = Y and Sc) have been investigated using quasielastic neutron scattering. 
The results reveal a localized motion on the ps time scale and with an activation energy of $\sim$10-30 meV, in both materials.
The temperature dependence of the total mean square displacement of the protons suggests an onset of this motion at a temperature of about 300 K.
Comparison of the QENS results to density functional theory calculations suggests that for both materials this motion can be ascribed to intra-octahedral proton transfers occurring close to a dopant atom.
The low activation energy, more than ten times lower than the activation energy for the macroscopic proton conductivity, suggests that this motion is not the rate-limiting process for the long-range proton diffusion, \textit{i.e.} it is not linked to the two materials significantly different proton conductivities.

\end{abstract}

\maketitle
\section{Introduction}

Hydrated acceptor-doped perovskites are known to be proton conductors in the temperature range $\sim$200--700$^\circ$C.\cite{KRE03} The doping creates an oxygen-deficient structure and in a humid atmosphere water molecules dissociate into hydroxide ions, which fill the oxygen vacancies, and protons, which bond to lattice oxygens.\cite{KRE03} The proton conductivity of perovskites varies quite dramatically amongst different types of materials for the same dopant concentration. For instance, in the BaZr$_{0.90}$$A$$_{0.10}$O$_{2.95}$ ($A$ = Y,  In, Sc and Ga) system, the proton conductivity differs several orders of magnitude depending on the dopant atom used.\cite{KRE01} 
The origin for this difference is intriguing and remains to be clarified.

The current understanding of the conduction mechanism in hydrated perovskites is that the transport process occurs in two elementary steps: i) hydrogen-bond mediated proton transfer between adjacent oxygens, and ii) reorientational motion of the hydroxyl group in between such transfers. In this two-stage mechanism it is only the protons that exhibit long-range diffusion while the oxygens reside on their crystallographic positions. These processes have been studied in different systems mainly using molecular dynamics simulations \cite{MUN97,KRE98,MUN96,SHI97} and quasielastic neutron scattering (QENS)\cite{HEM95,MAT96}. QENS is a particularly suitable experimental technique as it gives access to the time range $\sim$10$^{-13}$--10$^{-9}$ s, in which these processes occur, as well as providing information about the spatial geometry of the atomic motions. In addition, the very high neutron scattering cross section of protons provides a good contrast in the experiments and enables studies of systems with a low dopant concentration, \textit{i.e.} a low hydration degree and concomitantly few protons. 

Previous QENS investigations have revealed localized motions in the ps time range, mainly interpreted as rotational motions with radii between $r$$\sim$0.7 and 1.1 {\AA}, and with activation energies $E_{\rm{a}}$$\lesssim$100 meV.\cite{MAT96,PIO97} The results have also been interpreted in more complex terms of a diffusional process of the proton consisting of a sequence of free diffusion and trapping/escape events.\cite{HEM95} These studies have, however, been performed on orthorhombic perovskites, which makes the interpretation of the data more difficult as the structure is intrinsically anisotropic and there are several non-equivalent oxygen sites for the protons in the structure. In addition, no study has addressed the question of the role of the type of dopant on the elementary steps in the conduction mechanism.  

Molecular dynamics simulations on proton conducting perovskites have revealed a fast rotational motion of the protons around the line connecting the cell edge of the perovskite structure on the ps time scale and with an activation energy less than 50 meV.\cite{MUN97,KRE98,MUN96,SHI97} By extending the simulations to higher temperatures and pressures, also five proton transfer events within a simulation time of $\sim$100 ps were observed.\cite{KRE98} This process was found to be considerably dependent on the oxygen-oxygen separation, inferred by the vibrations of the oxygen sublattice.\cite{KRE98} In the contracted transition state, having an energy of $\sim$0.41 eV, proton transfer was found to occur almost barrierless, although it was not always that the proton was transferred.\cite{KRE98}

In the present work we investigate the proton dynamics in 10{\%} Y- and Sc-doped BaZrO$_{3}$, the two extremes in the cubic BaZr$_{0.90}$$A$$_{0.10}$O$_{2.95}$ ($A$ = Y, In, Sc and Ga) system considering the conductivity, using QENS. The main interest in this study lies in the fact that although the structure of these materials is very similar, their proton conductivities differ almost two orders of magnitude with the Y-doped perovskite exhibiting the highest.\cite{KRE01} The QENS results reveal a localized proton motion, with a slightly different activation energy in the two materials, and a comparison to density functional theory calculations suggests that for both materials we observe proton transfers between neighboring oxygens close to dopant atoms.

\section{Experimental details}

The BaZr$_{0.90}$$A$$_{0.10}$O$_{2.95}$ ($A$ = Y and Sc) samples, hereafter abbreviated as 10Y:BZO and 10Sc:BZO, respectively, were prepared by mixing stoichiometric amounts of BaCO$_{3}$, ZrO$_{2}$ and (Y/Sc)$_{2}$O$_{3}$. 
The oxides were heated to 250$^{\circ}$C overnight to remove moisture prior to weighing. 
To ensure thorough mixing ethanol (99.5 {\%}) was added during the milling procedure, which was performed manually using an agate mortar and a pestle. 
The finely ground mixtures were fired at 1000$^{\circ}$C for 8 hours and subsequently ground and pelletized using a 13 mm diameter die under a pressure of 8 tons. The pellets were sintered at 1200$^{\circ}$C in air for 72 hours. After sintering, the pellets were reground, compacted, and refired at 1500$^{\circ}$C for 48 h. Finally, the pellets were finely reground to powders. 
The charging with protons was performed by annealing the powder samples at $\sim$300$^{\circ}$C under a flow (12 ml/min) of Ar saturated with water vapor at 76$^{\circ}$C for 10 days.
Our X-ray measurements revealed a cubic structure of both hydrated materials. 
One should note that 10Y:BZO has previously been reported as tetragonal.\cite{KRE01}
However, from our diffractogram we find no evidence for a tetragonal distortion.

The QENS experiments were performed at the time-of flight spectrometer IN6 at the Institut Laue Langevin in Grenoble (France), using neutrons with a wavelength of 5.1 {\AA}, yielding an energy resolution of 100 $\mu$eV (FWHM) and a $Q$-range 0.26--1.9 {\AA}$^{-1}$. The $Q$-dependent spectra were summed into 14 groups to increase the statistics. Spectra were measured at 2, 250, 300, 380, 430, 465 and 495 K for 10Y:BZO, and at 2, 380, 430, 465 and 495 K for 10Sc:BZO, and the integration time was four hours. The 2 K measurements were used as resolution functions in the data analysis. The samples were loaded in vacuum tight Al-containers, which were coated on the inside  with a 100 nm thick layer of Pt to avoid corrosion. The sample thickness was chosen to 6 mm. 
The sample containers were mounted in 135$^{\circ}$ geometry in a cryofurnace. A spectrum of the empty sample container was measured at 380 K and subtracted as a background from the sample spectra while a vanadium standard was used to correct for the detector efficiency.

\section{Theoretical background}

In a QENS experiment one obtains the dynamic structure factor, $S(Q,\omega)$, which gives the probability that an incident neutron is scattered by the sample with a momentum transfer $\hbar$$Q$ and an energy transfer $\hbar\omega$. 
In our analysis we have modeled $S(Q,\omega)$ with the following scattering function
\begin{equation} 
\label{scatteringfunction}
\begin{split}
S_{\rm{q}}(Q,\omega) =&~b(Q)\cdot\delta(\hbar \omega) + \frac{1}{\pi}\frac{a_{\rm{L}}(Q)\cdot\Delta\omega(Q)}{((\hbar \omega)^{2}+[\Delta\omega(Q)]^{2})} \\
&+c(Q)+d(Q)\cdot(\hbar \omega).
\end{split}
\end{equation}
Here, the first term describes the elastic scattering from those atoms that move too slow to be resolved in the experiment ($b(Q)$ is a $Q$-dependent constant and $\delta$($\hbar\omega$) is a delta function).
The second term, which appears as a broadening of the elastic peak, accounts for stochastic motions (translational and reorientational). This result in a small energy loss/gain of the incident neutrons. 
At the IN6 spectrometer motions in the time range $\sim$10$^{-13}$ to 10$^{-10}$ s can be studied.
Since only protons are expected to move on this time scale the quasielastic broadening can thus be directly related to proton motions, which facilitates the analysis. The quasielastic scattering is described by a Lorentzian with amplitude $a_{\rm{L}}(Q)$ and width $\Delta\omega(Q)$ (HWHM) while $c(Q)$ and $d(Q)$ are parameters of a sloping background. It should be noted that all fit parameters were free throughout the fitting procedure.

The measured scattering function, $S_{\rm{meas}}(Q,\omega)$, is a convolution of the real scattering function, $S(Q,\omega)$, and the resolution function of the instrument, $R(Q,\omega)$, \textit{i.e.}
\begin{equation} 
\label{convolution}
\begin{split}
S_{\rm{meas}}(Q,\omega) &= S(Q,\omega) \otimes R(Q,\omega).
\end{split}
\end{equation}
Thus, the modeled scattering function in Eq.~(\ref{scatteringfunction}) is convoluted with the resolution function in the data analysis. 

The $Q$-dependences of $\Delta\omega$ and $a_{\rm{L}}$ contain information about the relaxation time and spatial geometry of the proton dynamics.
For instance, for a long-range diffusional process and for sufficiently small $Q$-values, the quasielastic width follows a characteristic $Q^{2}$-dependence according to $\Delta\omega$~=~$\hbar$$D$$Q^2$, where $D$ is the diffusion constant.\cite{BEE88}
For a local process, \textit{e.g.} a rotational motion, $\Delta\omega$ is practically $Q$-independent and is related to the relaxation time $\tau$ through $\tau$~=~$\hbar$/$\Delta\omega$.\cite{BEE88}  
One should here note that before analyzing the $Q$-dependence of the quasielastic intensities, $a_{\rm{L}}$, these need to be corrected for the overall decrease in scattering intensity due to the Debye-Waller factor, that accounts for the harmonic (vibrational) motions. 
The total scattering can be written as
\begin{equation} 
\label{DebyeWaller}
\begin{split}
S(Q,\omega)&=e^{-<u^2>_{\rm{harm}}Q^2}S_{\rm{q}}(Q,\omega)
\end{split}
\end{equation}
where $\langle$$u^2\rangle$$_{\rm{harm}}$ is the harmonic mean square displacement of all atoms in the material.
Thus, also the mean square displacement is needed for a full analysis of the experimental data. Following a custom route in the analysis of QENS data, the total mean square displacement ${\langle}u^2\rangle$, can be calculated from the ratio of the elastic and total scattering intensity according to
\begin{equation} 
\label{DWfactor}
\begin{array}{c}
\frac{{S_{\rm{meas}}}(Q,\omega=0)}{  \int^{}_{}{ S_{\rm{meas}}(Q,\omega)} d\omega} \vspace{2mm} = e^{-\langle u^{2} \rangle Q^{2}} \\
\ln\bigg(\frac{{S_{\rm{meas}}}(Q,\omega=0)}{  \int^{}_{}{ S_{\rm{meas}}(Q,\omega)} d\omega}\bigg) = -\langle u^{2} \rangle Q^{2}.
\end{array}
\end{equation}

\section{Results}

\begin{figure}
\includegraphics[width=0.48\textwidth]{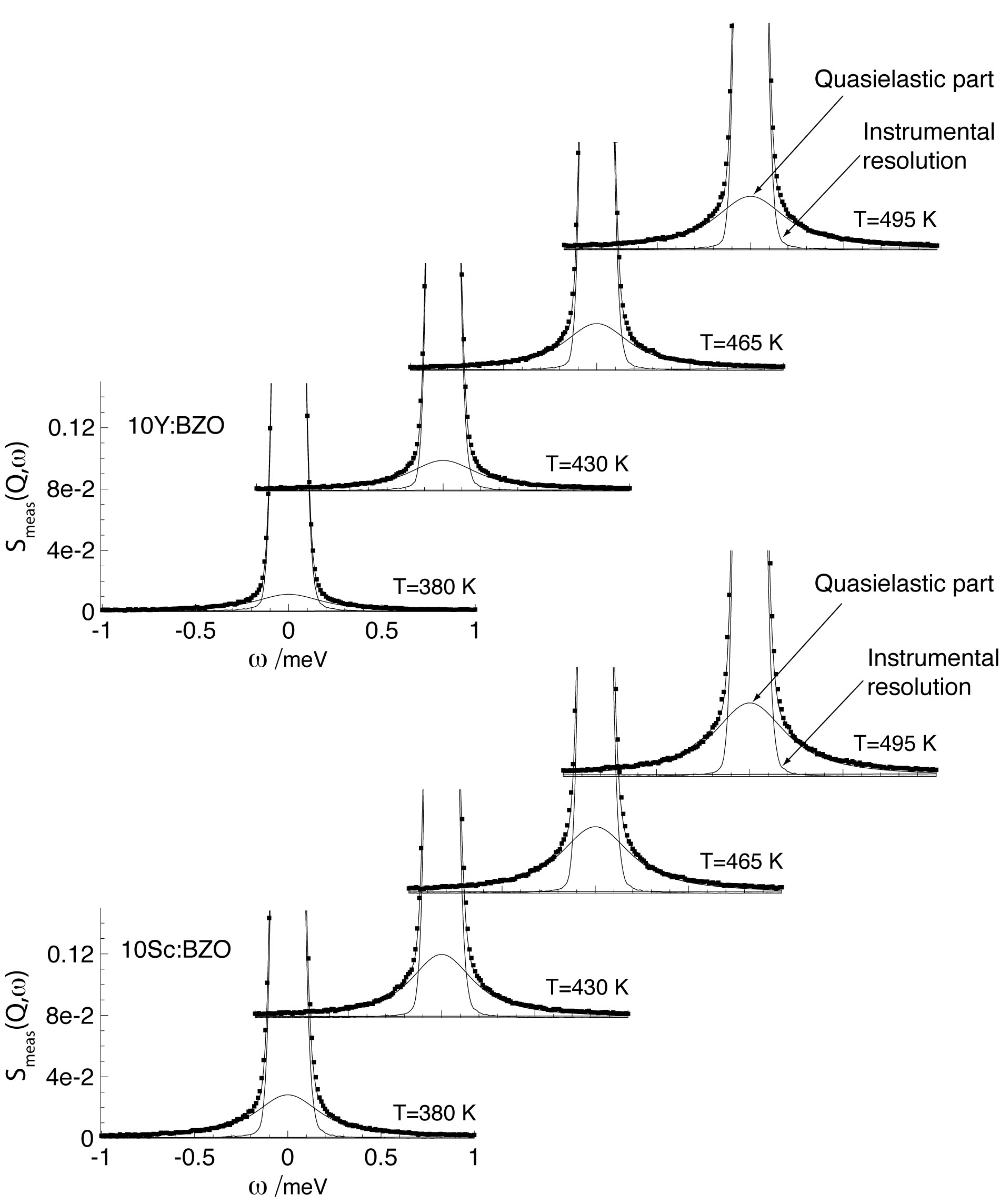}
\caption{\label{spectra} $S_{\rm{meas}}(Q,\omega)$ of 10Y:BZO and 10Sc:BZO at $Q$=1.9 {\AA}$^{-1}$ and for $T=380$--495 K. The solid lines represent fits to the experimental data (markers) according to Eq.~(\ref{scatteringfunction}). Each spectrum has been normalized to unity and cut at $S_{\rm{meas}}(Q,\omega)$~=~0.15 for easier comparison.
}
\end{figure}

Figure~\ref{spectra} shows the experimental data, $S_{\rm{meas}}(Q,\omega)$, for the two investigated materials for temperatures in the range 380-495 K, and for $Q$~=~1.9~{\AA}$^{-1}$.
As seen in the figure there is a quasielastic component together with a strong elastic component for both materials in each spectrum.
The quasielastic component increases with increasing temperature.

\begin{figure}
\includegraphics[width=0.37\textwidth]{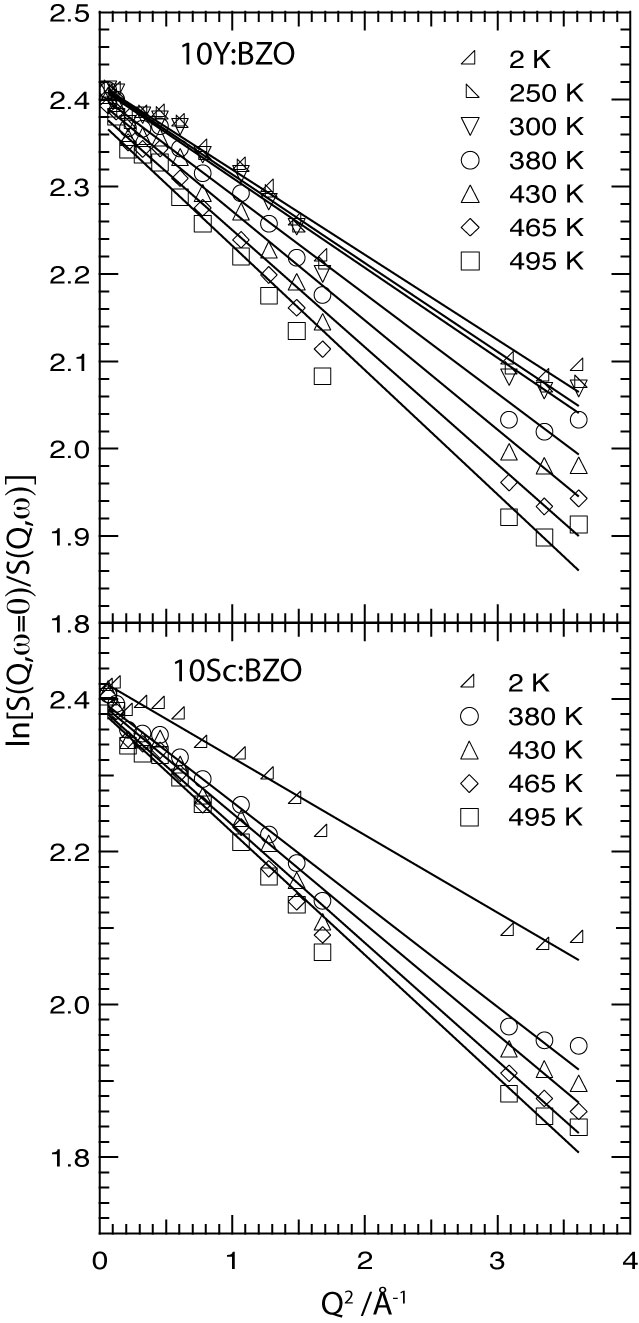}
\caption{\label{u2} $Q^2$-dependence of the logarithm of the elastic intensity normalized to the total intensity, where the error bars of the data points are within the size of the symbols. 
The lines represent linear fits with a slope -${\langle}u^2\rangle$. 
}
\end{figure}

\begin{figure}
\includegraphics[width=0.37\textwidth]{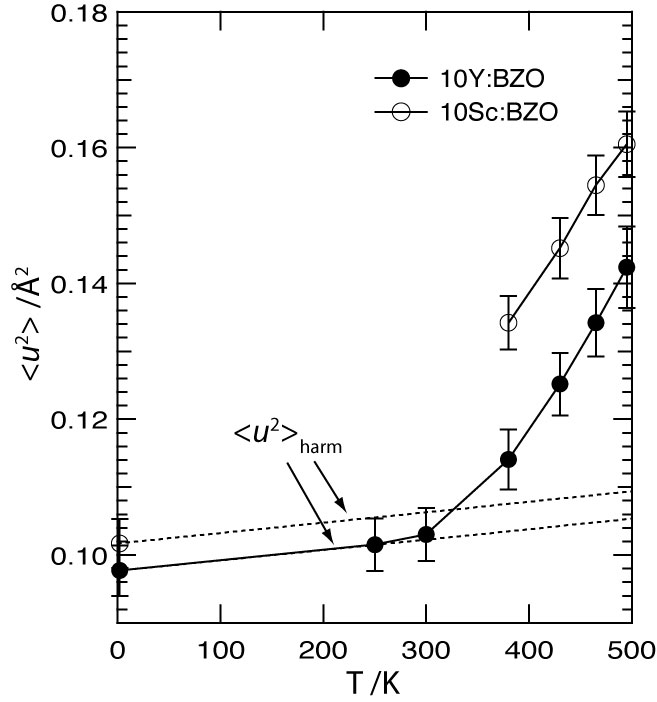}
\caption{\label{u2T} Temperature dependence of the total mean square displacement ${\langle}u^2\rangle$. The dashed lines represent the estimated harmonic mean square displacement, ${\langle}u^2\rangle$$_{\rm{harm}}$, for the two materials.}
\end{figure}

To determine when we can actually expect to observe the stochastic motion of the protons in the experiment we have calculated the total mean square displacement, ${\langle}u^2\rangle$, using Eq.~(\ref{DWfactor}). 
Figure~\ref{u2} shows the logarithm of the ratio of the elastic and total intensity versus $Q^2$, together with linear fits from which ${\langle}u^2\rangle$ has been determined.
We have discarded the spectra for 1.35 {\AA}$^{-1}$ $<$ $Q$ $<$ 1.75 {\AA}$^{-1}$ due to a Bragg reflection resulting in a strong elastic intensity masking the quasielastic contribution. 
The total intensity is approximated by integrating the spectrum over the region -2~$<$~$\omega$~$<$~2 meV.
Figure~\ref{u2T} shows the temperature dependence of the so obtained ${\langle}u^2\rangle$.
For 10Sc:BZO we have only performed experiments at 2 K and $T\geq380$~K, whereas for 10Y:BZO, we have also data for two temperatures in between. 
For 10Y:BZO at low temperatures, $T<300$~K, the mean square displacement increases only slightly and basically linearly with temperature, while above 300 K the temperature dependence is much stronger. 
The linear increase at low temperature reflects an increasing amplitude of harmonic vibrations while the non-linear increase above 300 K results from that we also resolve proton motions other than vibrational. 
The same behaviour can be inferred for 10Sc:BZO, even though we here do not observe the transition in the mean square displacement from a vibrational to a stochastic motion due to the lack of data in this region. 
Over the whole temperature range, the mean square displacement is slightly larger in the Sc-doped material compared to the Y-doped counterpart.

\begin{figure}
\includegraphics[width=0.37\textwidth]{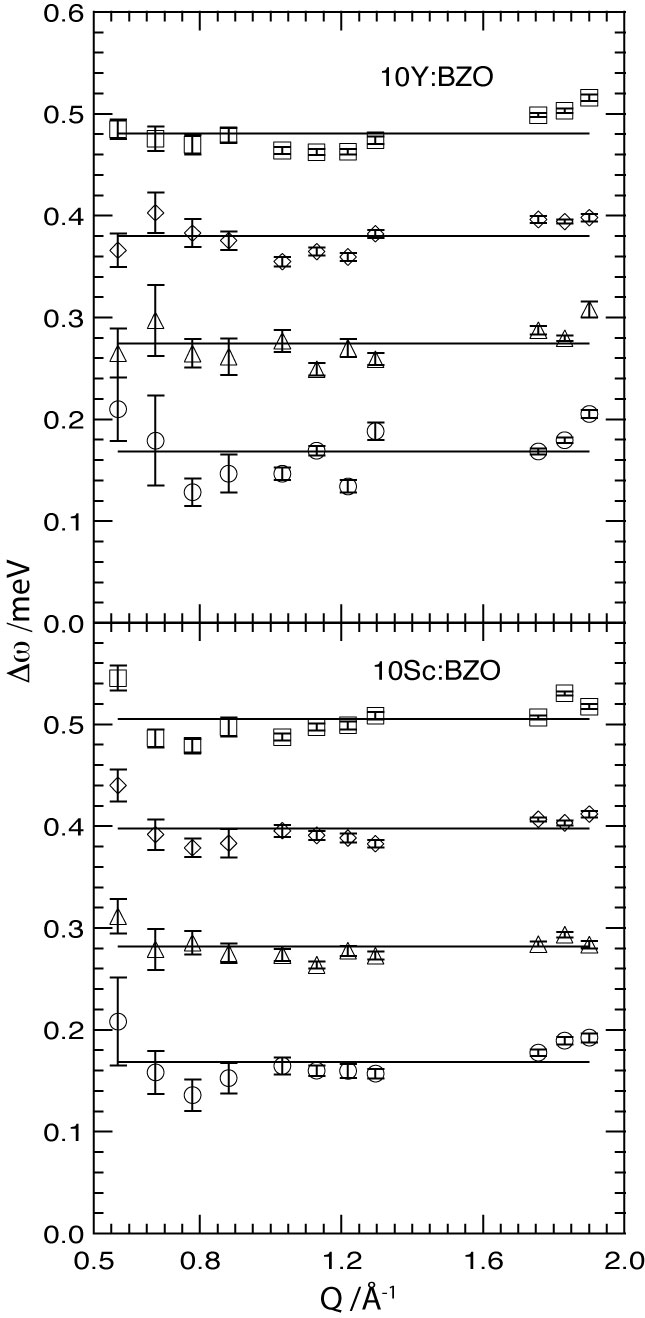}
\caption{\label{wL} 
Plots of the quasielastic widths and corresponding linear fits at 380 K ($\circ$), 430 K ($\triangle$), 465 K ($\Diamond$) and 495 K ($\square$).
The data points and fits have been vertically shifted by 0.1 meV (430 K), 0.2 meV (465 K) and 0.3 meV (495 K) for clarity.
}
\end{figure}

\begin{figure}
\includegraphics[width=0.37\textwidth]{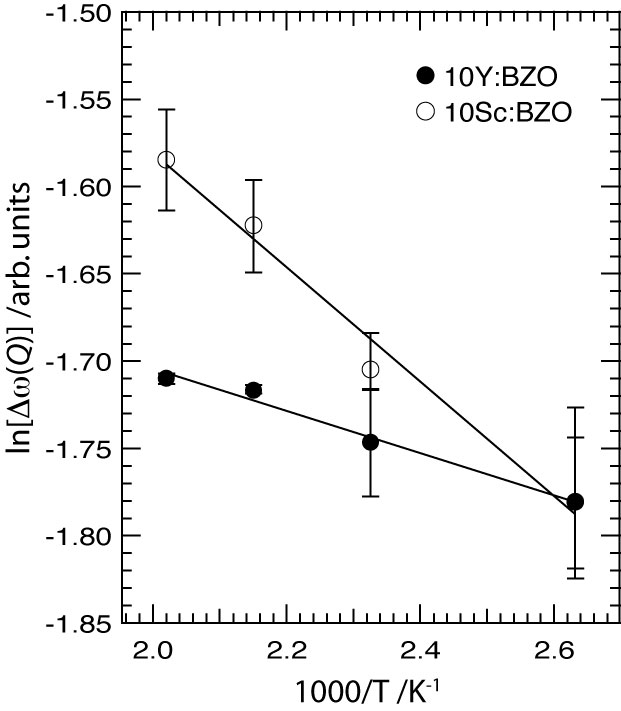}
\caption{\label{Ea_wL} 
Arrhenius plot of the quasielastic widths with linear fits which correspond to activation energies of 10 meV (10Y:BZO) and 30 meV (10Sc:BZO), respectively.
} 
\end{figure}

Consistent with the mean square displacement we indeed observe a quasielastic broadening for $T\ge300$ K. 
Figure~\ref{wL} shows the quasielastic widths as a function of $Q$, obtained from the fits of Eq.~(\ref{scatteringfunction}) to the spectra. 
As seen, the widths are practically $Q$-independent and have values in the range 0.17--0.21 meV.
There is a slight temperature dependence of the quasielastic widths. In order to extract this dependence we have fitted the largely $Q$-independent widths by a constant at each temperature, represented by the lines in Fig.~\ref{wL}. An Arrhenius plot of the so obtained quasielastic widths are shown in Fig.~\ref{Ea_wL}. 
From this figure we obtain an activation energy of approximately 10 meV for 10Y:BZO and 30 meV for 10Sc:BZO. 
As also seen in the figure, the width is larger for the Sc-doped material for all temperatures except for $T=380$~K where they are the same.

\begin{figure}
\includegraphics[width=0.37\textwidth]{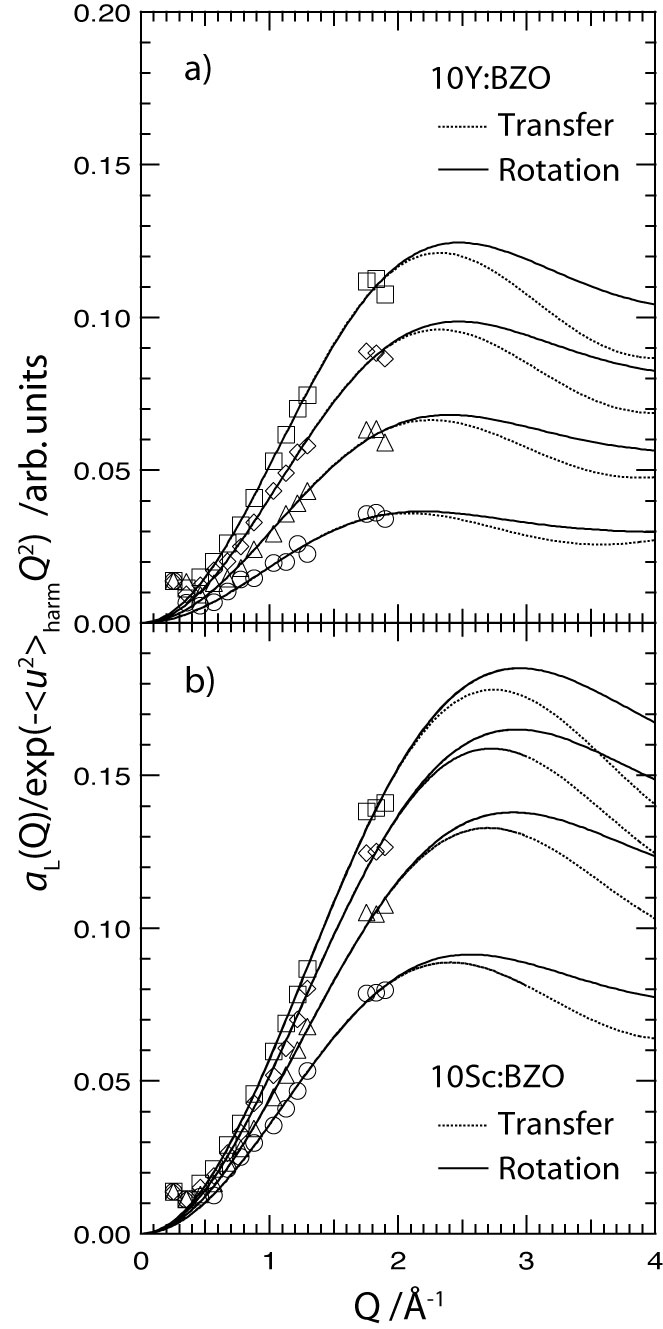}
\caption{\label{aL} 
$Q$-dependence of the quasielastic intensities at 380 K ($\circ$), 430 K ($\triangle$), 465 K ($\Diamond$) and 495 K ($\square$), respectively.
Solid lines are fits to the data using a jump-diffusion model over two (transfer) and four (rotation) sites, according to Eq.~(\ref{structurefactor_transfer}) and Eq.~(\ref{structurefactor_rotation}). 
}
\end{figure}

Figure~\ref{aL} shows the Debye-Waller corrected quasielastic intensities versus $Q$ for $T$~=~380--495 K, for the two materials.
It should be noted that the total mean square displacement in Fig.~\ref{u2T} cannot be used directly to calculate the Debye-Waller factor for these temperatures, as this should only contain the harmonic, vibrational, contribution.
Therefore, for $T$~=~380--495 K we have calculated the Debye-Waller factor from a linear extrapolation of the vibrational mean square displacement between 2 and 250 K for the 10Y:BZO material. 
For 10Sc:BZO, we have estimated the harmonic mean square displacement by using the same slope as found for 10Y:BZO and an extrapolation from the 2K value, see Fig.~\ref{u2T}.
As seen in Fig.~\ref{aL}, the quasielastic intensity increases with increasing temperature and with $Q$. 
We find that the quasielastic intensity is slightly higher for the Sc-doped material.

\section{Discussion}

The $Q$-independence of the quasielastic width observed in our experiment for both 10Y:BZO and 10Sc:BZO suggests that we are following a local proton motion.\cite{BEE88} 
Based on the values of the widths (0.17--0.21 meV) this motion has a relaxation time in the range 3.1--3.9 ps, depending on the temperature.
The temperature dependence of the total mean square displacement in Fig.~\ref{u2T} suggests an onset of this motion at a temperature of about 300 K, whereas below this ''threshold'' temperature, the total mean square displacement seems to be related only to harmonic (vibrational) motions of the protons.
This finding agrees well with a recent inelastic neutron scattering study of hydrated BaIn$_{0.20}$Zr$_{0.80}$O$_{2.90}$, where it was found that the total mean square displacement of the protons increased only very slightly between 30 and 300 K.\cite{KAR07_Tosca} 

The observed local process has an activation energy of 10 meV (10Y:BZO) and 30 meV (10Sc:BZO) for the quasielastic width.
These values are considerably lower than the activation energy for the proton conductivity, which is around 500 meV.\cite{KRE01} 
Thus, the process we observe in our experiment cannot be rate limiting for the long-range proton diffusion. 
That is, there must exist another dynamical process of the protons, which occurs on a longer time scale and that cannot be observed in the present experiments, that controls the rate of the long-range diffusivity, \textit{i.e.} the macroscopic proton conductivity.
Perhaps, this could be the escape event from a trapping site in the structure, where the proton spends an extended time before it diffuses further.
Furthermore, we note that the observed proton motion is slightly faster in 10Sc:BZO and that the quasielastic scattering intensity is slightly higher. 
Considering that 10Y:BZO has higher conductivity than 10Sc:BZO and a higher degree of hydration,\cite{KRE01} \textit{i.e.} more protons in the structure, also point towards that we are observing a process that does not determine the rate of the long-range proton diffusion.

As described above, we can imagine two localized dynamical processes for the proton in hydrated perovskites, i) proton transfer between adjacent oxygens, and ii) rotation of the O-H group around the line connecting the octahedrally coordinated cations. 
Density functional theory (DFT) calculations have shown that these processes are well described with a jump diffusion model over two (transfer) and four (rotation) sites, respectively.\cite{BJO05}
For these models the quasielastic intensities are given by \cite{BEE88}
\begin{equation}
\label{structurefactor_transfer}
a_{\rm{L,2}}(Q) =  \alpha\cdot\Big\{  \frac{1}{2}-\frac{1}{2}\frac{\sin(2Qr)}{2Qr} \Big\}
\end{equation}
and
\begin{equation}
\label{structurefactor_rotation}
a_{\rm{L,4}}(Q) =  \alpha\cdot\Big\{\frac{3}{4}-\frac{1}{4}\frac{\sin(2Qr)}{2Qr}-\frac{1}{2}\frac{\sin(\sqrt{2}Qr)}{\sqrt{2}Qr} \Big\},
\end{equation}
where $a_{\rm{L,2}}(Q)$ and $a_{\rm{L,4}}(Q)$ are the quasielastic structure factors for the transfer and rotational motion, respectively, $r$ is the jump length, and $\alpha$ is a scale factor that we need to use since we do not have the quasielastic intensities on an absolute scale.
Included in Fig.~\ref{aL} are free fits with the two model functions to the quasielastic intensity for the investigated temperatures, while the fit parameters are shown in Tab.~\ref{fitparametersdecoupled}.
As can be seen in Fig.~\ref{aL}, it is very difficult to discriminate between the two models.
In order to directly separate them it would be necessary to have data at larger $Q$-values. 
It should here be noted that some of us performed an experiment at the inverted time-of-flight quasielastic spectrometer IRIS at the pulsed neutron source ISIS in United Kingdom to extend the $Q$-range to 0.5--3.7 {\AA}, but due to the presence of numerous Bragg peaks in this $Q$-range it was, unfortunately, impossible to analyze these spectra with sufficient accuracy.

Insight into what process we are observing may instead be obtained from the jump lengths, see Tab.~\ref{fitparametersdecoupled}.
From these we can see that for both the transfer and rotational model the jump lengths for the Sc-doped material are overall shorter than for the Y-doped counterpart. 
Furthermore, we see that apart from the lowest temperature (380 K), the obtained jump lengths are basically temperature independent.
The temperature independence for $T=430$--495 K suggests that the fits are physically reasonable, while the overall larger values for the 380 K data is likely a consequence of the low quasielastic intensity at this temperature, wherefore these fit parameters have larger uncertainty. Thus, we will keep the following discussion to $T\geqslant430$~K.

\begin{figure}
\includegraphics[width=0.25\textwidth]{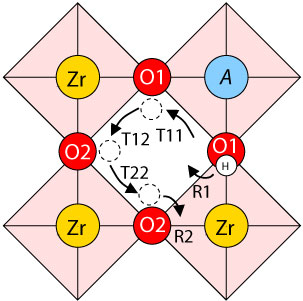}
\caption{\label{picture} 
Schematic picture of the migration paths for a proton in the vicinity of a dopant atom $A$, redrawn from Ref.\cite{BJO07_2}.
}
\end{figure}

The proton experiences a different local surrounding and hence a different potential energy surface (PES) depending on where it is located in the structure.\cite{KAR05,BJO07_2}
Figure~\ref{picture} shows the relevant migration paths for the proton around and between oxygens in the first (O1) and second (O2) oxygen coordination shell of a dopant atom $A$.
As seen in Fig.~\ref{picture}, R1 and R2 denote the rotational processes at O1 and O2, respectively, and T11, T12 and T22 represent proton transfers between adjacent O1 atoms, between adjacent O1 and O2 atoms, and between adjacent O2 atoms, respectively. Further away from the dopant atom the PES is less distorted and there is essentially only one type of rotational (R) and one type of transfer (T) process. These local motions have been modeled by DFT calculations on a system corresponding to a dopant concentration of 3.7{\%}.\cite{BJO07_2} The jump lengths and classical migration barriers so obtained for the R1, R2, T11, T12, T22, R and T processes in Sc- and Y-doped BaZrO$_{3}$, respectively, are reported in Tab.~\ref{DFT}.
\begin{table}
\caption{\label{fitparametersdecoupled}
Jump lengths obtained from free fits of the quasielastic intensity with the jump-diffusion models over two and four sites according to Eq.~(\ref{structurefactor_transfer}) and Eq.~(\ref{structurefactor_rotation}).
$r_{2}$ denotes the jump length for the transfer process, and $r_{4}$ denotes the jump length (radius) for the rotational process.
}
\begin{tabular}{| l | cc | cc |}
\hline
&  10Sc:BZO & & 10Y:BZO & \\
T/K & $r_{2}$/{\AA}& $r_{4}$/{\AA}
& r$_{\rm{2}}$/{\AA} & r$_{\rm{4}}$/{\AA} \\
\hline
380     & 0.93$\pm$0.14 & 1.10$\pm$0.08 & 1.08$\pm$0.18 & 1.29$\pm$0.12 \\ 
430     & 0.83$\pm$0.10 & 0.98$\pm$0.06  & 0.99$\pm$0.12 & 1.18$\pm$0.10 \\ 
465     & 0.82$\pm$0.12  &  0.97$\pm$0.08 & 0.97$\pm$0.12 & 1.15$\pm$0.08 \\ 
495     & 0.82$\pm$0.10 & 0.96$\pm$0.08 & 0.97$\pm$0.12 & 1.14$\pm$0.08 \\ 
\hline
\end{tabular}
\end{table}
By comparing these jump lengths to the experimental results in Tab.~\ref{fitparametersdecoupled}, and by also taking the structure of the PES into account, we may draw conclusions about which process/es we are observing in the experiment.

For the Sc-doped material the DFT calculations\cite{BJO07_2} reveal that the PES is much lower at and around O1, \textit{i.e.} in the vicinity of the Sc atoms, than at O2 or at an O located even further away from the dopant. 
This means that the probability for finding a proton is higher close to a dopant than farther away, which in turn suggests that the local process observed in the experiment is either R1 or T11. 
The experimental jump length obtained from fitting with a transfer model, $r_2$, is 0.82~{\AA} (\textit{cf.} Tab.~\ref{fitparametersdecoupled}), which is about 0.1~{\AA} longer than the DFT value for T11, given in Tab.~\ref{DFT}. 
The jump length (radius) obtained from fitting with the rotational model, $r_4$~=~0.97~{\AA}, is on the other hand in excellent agreement with the DFT value for R1. 
From a structural point of view this would suggest that the observed local proton motion in 10Sc:BZO is the rotational motion around an oxygen close to a Sc atom. 
However, this conclusion is not supported by the classical migration barriers (activation energies) presented in Tab.~\ref{DFT}. The R1 process has a classical barrier of 210~meV, which is much higher than the 30~meV observed in the QENS measurements. 
Correcting for the difference in zero point energy, $\Delta \mathrm{ZPE}$, of the proton at the stable site and at the barrier top is expected to reduce the calculated barrier with approximately 40~meV.\cite{SUN07}
Consequently, this would still give a barrier of 170~meV, wherefore it is unlikely that R1 is the process we observe in the QENS experiments. 

For T11 the DFT calculations give a much lower classical energy barrier, only 130~meV compared to 210~meV for R1. 
By further adding the zero point contribution ($\Delta \mathrm{ZPE} \approx -120$~meV for proton transfers in BZO\cite{SUN07}) the theoretical barrier for T11 is substantially reduced to approximately 10~meV, which is close to the energy barrier determined from the experiments (30~meV).
Based on this it is therefore more likely that it is the T11 process that is observed.
It should be noted, though, that the theoretical estimate most probably is an underestimation of the real T11 barrier, as DFT at the level of complexity used in Ref.\cite{BJO07_2} has a tendency to give too small barriers for hydrogen-bond mediated processes.\cite{Barone1996} 
Admittedly there is a discrepancy between the experimental and theoretical jump lengths, see Tab.~\ref{fitparametersdecoupled} and Tab.~\ref{DFT}, but not so large that it cannot be explained by the errors inherent in the two methods. 
First of all, the experimental error bar is $\sim$0.1~{\AA}.
Secondly, there is also an uncertainty in the calculated value. 
The ability of standard DFT approaches to accurately predict hydrogen bond lengths depends sensitively on the local geometry of the hydrogen bond.\cite{Ireta2004} 
Often, but not always, the hydrogen bond length is underestimated, sometimes with as much as $\sim 0.1$~{\AA}.\cite{Barone1996,Ireta2004}      
\begin{table}
\caption{\label{DFT}
Jump lengths, $r$, and classical energy barriers (activation energies), $E_\mathrm{a}$, obtained from DFT calculations for the different migration paths in doped BZO.\cite{BJO07_2,BJO_unpublished} R1, $\ldots$, R2 are the paths in the near dopant region, as depicted in Fig.~\ref{picture}, whereas R and T are the rotation and transfer paths, respectively, far from dopant atoms.
}
\begin{tabular}{| l | cc | cc |}
\hline
Migration &  3.7Sc:BZO & & 3.7Y:BZO & \\
path & $r$/{\AA}& $E_\mathrm{a}$/meV & $r$/{\AA}& $E_\mathrm{a}$/meV \\
\hline
R1   & 0.96 & 210 & 0.97 & 260 \\
T11 & 0.72 & 130 & 0.80 & 90 \\
T12 & 0.89 & 570 & 0.79 & 280 \\
T22 & 0.89 & 210 & 1.04 & 340 \\
R2  & 0.89 & 220 & 0.89 & 320 \\
\hline
R   & 0.98 & 200 & 0.98 & 200 \\
T   & 0.82 & 170 & 0.82 & 180 \\
\hline
\end{tabular}
\end{table}

For the Y-doped material the situation seems to be very similar although for this material the PES is not only lowered at O1 but also, to an equal extent, at O2.\cite{BJO07_2}
This suggests that the protons spend most of their time close to the Y dopants, but, in contrast to Sc:BZO, with a similar occupation of O1 and O2 sites. 
However, inspection of Tab.~\ref{DFT} shows that all the migration pathways R1, $\ldots$, R2 near Y, except T11, are associated with relatively high classical energy barriers ($\sim 300$~meV). Hence, these processes are expected to occur on much longer time scales than the one/ones we observe in the experiments. The only remaining candidate is then T11, which has a calculated classical barrier is of 90~meV (\textit{cf.} Tab.~\ref{DFT}). 
 Further addition of $\Delta \mathrm{ZPE} \approx -120$~meV obviously results in a negative barrier, which is not physically sound. Given the tendency of DFT to underestimate this type of barrier it is, however, not unreasonable to link T11 to the measured 10~meV activation energy in the Y-doped material.   
Consequently, we once again claim that T11 gives the predominant contribution to the experimental spectra, despite an even larger (0.17~{\AA}) difference between the experimental and theoretical jump lengths.
In this context we note that even though the degree of hydration is higher in the Y-doped material,\cite{KRE01} the slightly higher quasielastic intensity for the Sc-doped material (\textit{c.f.} Fig.~\ref{aL}) suggests that there is a larger number of protons in the immediate vicinity of the dopant in the Sc-doped material than in the Y-doped counterpart.
This is in agreement with the very high T12 barrier for the Sc-doped material, which may also be the reason to the lower proton conductivity of this material.

Finally, it should be noted that the comparatively larger disagreement between our experimental results and the calculations in 10Y:BZO may suggest that a more complex process is taking place, or that the processes listed in Tab.~\ref{DFT} contribute in an averaged manner to the quasielastic intensity. 
In order to elucidate what process/es we are seeing in the two materials, and in 10Y:BZO in particular, further experimental and theoretical investigations are needed.

\section{Conclusions}

Proton motions in the hydrated proton conducting perovskites BaZr$_{0.9}$$A$$_{0.10}$O$_{2.95}$ ($A$ = Y and Sc) have been investigated using quasielastic neutron scattering. 
The results reveal a localized proton motion on the ps time scale and with a low activation energy, $\sim$10--30 meV.
Comparison of the QENS results to density functional theory calculations suggests that for both materials we observe proton transfers between neighboring oxygens bound to a dopant atom. 
The low activation energy, much lower than the activation energy for the macroscopic proton conductivity, suggests that this process is not rate-limiting for the long-range proton diffusion in either of the two materials.
That is, there must exist another dynamical process of the protons, which occurs on a slower time scale and that cannot be observed in the present experiment, that controls the rate of the long-range diffusivity, perhaps the escape event from a trap, where the proton spends a longer time before it diffuses further.
So, even though the two investigated materials possess considerably different proton conductivities, this difference cannot be directly linked to the observed proton motion.

\acknowledgments
This work was supported by the Research Council, the National Graduate Schools in Scientific Computing and Material Science and the Foundation for Strategic Research via the ATOMICS program, Sweden. Allocations of computer resources through the Swedish National Allocation Committee are gratefully acknowledged.

%\bibliography{../../../../References}
\end{document}